\documentclass[9pt]{article}
\usepackage{subeqnarray,subfigure}
\usepackage{epsfig,amsmath,amssymb,mathtools}
\usepackage{dirtytalk,enumitem}
\usepackage{mathrsfs,authblk}
\usepackage{tikz,lipsum,lmodern}
\usepackage[most]{tcolorbox}

\usepackage{empheq}
\usepackage{soul,ulem}
\usepackage[pagebackref=true, colorlinks=true]{hyperref}
\definecolor{redish}{rgb}{0.7,0.2,0.0}  
\definecolor{bluish}{rgb}{0.2,0.5,0.8}
\hypersetup{linkcolor=redish,          
    citecolor=blue,        
    filecolor=magenta,      
    urlcolor=bluish}          
\usepackage{setspace}

\title{Analysing the time period of Vela pulsar}
\author[1]{Shreyan Goswami\footnote{goswamishreyan@gmail.com}}
\author[2]{Hershini Gadaria \footnote{i19ph041@phy.svnit.ac.in}}
\author[3]{Sreejita Das \footnote{i19ph017@phy.svnit.ac.in}}
\author[4]{Midhun Goutham \footnote{i19ph050@phy.svnit.ac.in}}
\author[5]{Kamlesh N. Pathak \footnote{knp@phy.svnit.ac.in}}
\affil[1,2,3,4,5]{{\small Sardar Vallabhbhai National Institute of Technology, Surat, India}}
\date{~}
\begin{document}
\maketitle
\begin{abstract}
  In this project, we have implemented our basic understanding of Pulsar Astronomy to calculate the Time Period of Vela Pulsar. Our choice of pulsar rests on the fact that it is the brightest object in the high energy gamma ray sky. The simplistic data set consisting of only voltage signals makes our preliminary attempt as closely accurate as possible. The observations had been made at 326.5 MHz through a cylindrically paraboloid telescope at Ooty. A higher frequency creates a much lower delay in the arrival time of pulses and makes our calculations even more accurate. Being an already widely studied celestial body, it gives us the opportunity to compare our findings and make necessary modifications.
\end{abstract}
\section{Introduction}
Pulsars are rapidly rotating highly magnetised neutron stars. They emit two steady, narrow beams of electromagnetic radiation in opposite directions that sweep the sky like a lighthouse. Pulsars are of extreme importance to astronomers as they can help locate planets or other celestial bodies orbiting around it, measure the distance to galaxies, construct models of free electron distribution and detect gravitational waves. Calculating several parameters of known pulsars like its Distance or Time Period can allow us to perform further complex calculations and estimations, helping us gain a much deeper understanding of the universe. We have been provided with a raw voltage signal from the observation of the Vela Pulsar $(PSR B0833-45)$ by the two sub-apertures (north and south) of the Ooty Radio Telescope \cite{4}. It is a cylindrical paraboloid telescope based on a north-south slope of 11.2 degrees in Ooty. The reflecting surface is 530m long and 30m wide and is operated at 326.5 MHz. The large reflecting surface makes the telescope highly sensitive. The observations, as recorded in the data set, have been made at 326.5 MHz with a bandwidth of 16.5 MHz. A data set with one second’s worth of data was used with each row of data being separated by 30 nano seconds (\textcolor{brown} {33.3(3)}  MHz). The main difference in the analysis being, we have used the DM (Dispersion Measure) that has already been found to make sure our results have increased accuracy.\\ \\
In section \ref{section:2} we have explored the statistical characteristic of the \textcolor{brown} {Vela pulsar} to make sure there are no discrepancies in data. Voltage and power signals are plotted to understand how the distribution of the data is.\\ \\
Section \ref{section:3} discusses properties of the signal and dynamic spectrum. This section discusses the RFI and how one would eliminate it. The frequency delay results from the same.\\ \\
Section \ref{section:4} we find the distance using the correct values of DM to show that the same shall be applied to find a more accurate Time Period that allows us to eliminate the time delays by using the DM and get a dedispersed time series of the pulsar. In further sections, we find the average time period and plot the average profile of the pulsar.
\section{Statistical Characteristics of the Signal}\label{section:2}
We began our analysis by performing statistical evaluation of the raw voltage signals to verify some of its expected properties. We expect the signals to have a Gaussian distribution. To do this, we randomly and uniformly selected 100,000 voltage samples from both the north and the south arrays and plotted the histogram. As expected, the voltage signals demonstrate Gaussian distribution.
\subsection{Voltage Signal Characteristics}
We began our analysis by performing statistical evaluation of the raw voltage signals to verify some of its expected properties. We expect the signals to have a Gaussian distribution. To do this, we randomly and uniformly selected 100,000 voltage samples from both the north and the south arrays and plotted the histogram. As expected, the voltage signals demonstrate Gaussian distribution.
\begin{figure}[!htb]
    \centering
    \begin{minipage}{.5\textwidth}
        \centering
        \includegraphics[width=0.7\linewidth, height=0.15\textheight]{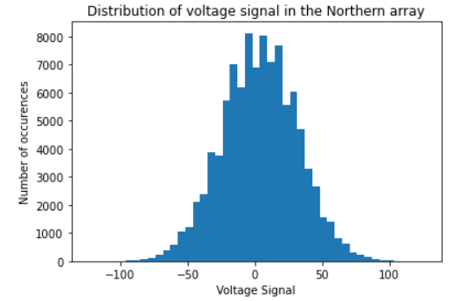}
        \caption{Histogram for northern array}
        
    \end{minipage}%
    \begin{minipage}{0.5\textwidth}
        \centering
        \includegraphics[width=0.7\linewidth, height=0.15\textheight]{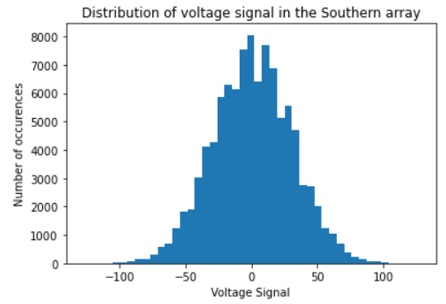}
        \caption{Histogram for southern array}
        
    \end{minipage}
    \caption{Voltage signal distribution of 100,000 randomly selected samples}    
\label{fig:voltagechar}
\end{figure}
\subsection{Power Signal Characteristics}
We further studied the distribution that the power signals follow. The power or intensity signal is merely the square of the voltage signal. The power signals are expected to follow an exponential distribution. We used the same data samples that we used to look at voltage characteristics previously. Histograms for both the northern and southern arrays were plotted. As expected, the power signals demonstrate an exponential distribution. 
\begin{figure}[!htb]
    \centering
    \begin{minipage}{.5\textwidth}
        \centering
        \includegraphics[width=0.7\linewidth, height=0.15\textheight]{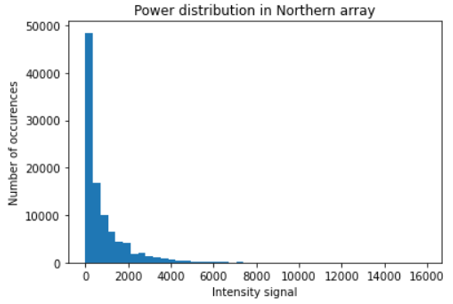}
        \caption{Histogram for northern array}
        
    \end{minipage}%
    \begin{minipage}{0.5\textwidth}
        \centering
        \includegraphics[width=0.7\linewidth, height=0.15\textheight]{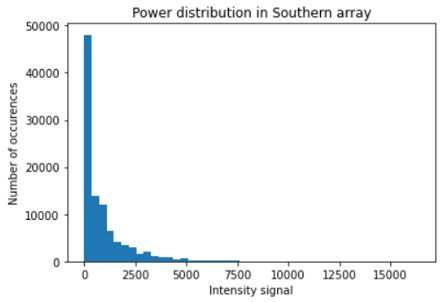}
        \caption{Histogram for southern array}
        
    \end{minipage}
\caption{Power signal distribution of 100,000 randomly selected samples}    
\label{fig:powerchar}
\end{figure}
\section{Properties of the signal in the Time-Frequency domain}\label{section:3}
\subsection{Voltage power spectrum}
\begin{figure}[!htb]
    \centering
    \begin{minipage}{.5\textwidth}
        \centering
        \includegraphics[width=0.7\linewidth, height=0.15\textheight]{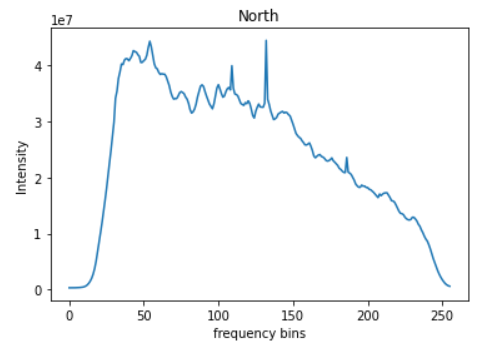}
      \caption{Voltage power spectrum for northern array}
        
    \end{minipage}%
    \begin{minipage}{0.5\textwidth}
        \centering
        \includegraphics[width=0.7\linewidth, height=0.15\textheight]{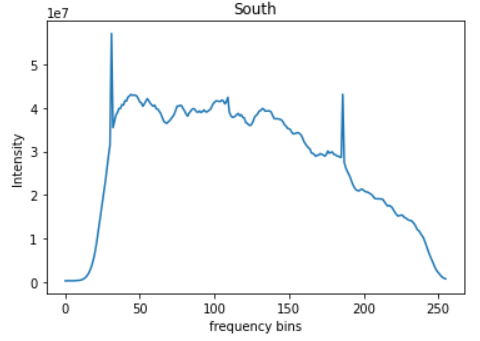}
        \caption{Voltage power spectrum for southern array}
        
    \end{minipage}
\caption{Average voltage power spectrum for both the arrays. The frequency axis ranges from 0 to 256 MHz}    
\label{fig:voltagepower}
\end{figure}

The Power Spectrum of a signal describes the power present in the signal as a function of frequency. Any physical signal can be decomposed into a spectrum of frequencies over a range.  We used a very efficient algorithm known as the Fast Fourier Transform (FFT) to plot the power distribution as a function of frequency. To compute the FFT, 256 frequency channels were used since power of 2 increases the speed of FFT (in this case, $16^2$). The average power spectrum of the voltage signal was plotted by averaging the power spectrum obtained from all 512-point FFTs. The voltage power spectrum for both the northern and southern arrays are given in figure \ref{fig:voltagepower}. Each Power Spectrum corresponds to an interval of 512/33 microseconds. The sharp peaks in the spectra might indicate the presence of local Radio Frequency Interference (RFI). RFI is a disturbance caused by an external source like cellular networks, lightning, solar flares, etc that affects the electrical circuit used to originally measure the voltage signals from the pulsar.
 As observed, the DC channel power is much larger in the northern array than in the southern. The plot smoothly tapers off to 0 at both edges, indicating that the aliasing is minimal. 
\subsection{Dynamic Spectrum}
The Dynamic Spectra is a color-coded graph that shows the relationship between Frequency (MHz) and Time (ms). It enables us to detect pulsar signal indicators. Incoherent Addition was used to combine the power from the two halves of the array to increase the Signal to Noise Ratio (SNR). Incoherent addition helped in removing Radio Frequency Interference (RFI) to a large extent. The dynamic spectrum is shown below. The x-axis represents time measured in ms and the y axis represents frequency in MHz. The colour bar on the right indicates the intensity of the power signal for a particular time and frequency data point. 
\begin{figure}[ht]
    \centering
    \includegraphics{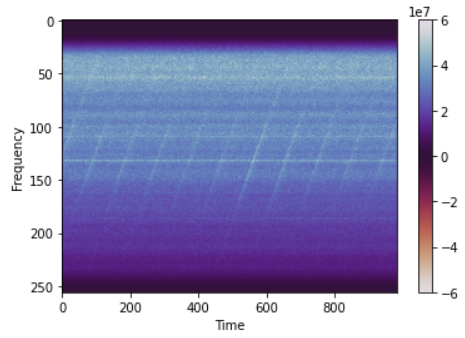}
    \caption{The dynamic spectrum of the signal. Frequency (in Mhz) is plotted on the y-axis, and decreases upwards. Time is plotted on the x-axis for a duration of 1000 ms.}
    \label{fig:dynamicspectrum}
\end{figure}
The diagonal and uniformly spaced features shown in the graph above leads us to conclude that the source has to be a pulsar. Upon carefully observing each pulse, the signal appears first at higher frequencies, and gradually appears later at lower frequencies. Thus, there is a frequency delay in the observed data which is a characteristic sign of a signal dispersed in the interstellar medium. Since our analysis is concerned with magnitudes, negative intensities at either ends of the colour bar do not pose any discrepancy.
\subsection{Dispersion Measure and the frequency delay}
The Dispersion Measure (DM) is a parameter that shows up in observations as the broadening of an otherwise sharp pulse. In statistics, it refers to how far a distribution may be stretched or squeezed. The DM is measured in pc/cc and is calculated as:
\begin{equation} \label{eq:1}
    t \approx t_\infty + 4.149 \times 10^3 \times DM \times \nu ^{-2}
\end{equation}
where $t$ is the pulse arrival time in seconds and $t_\infty$ is the pulse arrival time in seconds at infinite frequency. The DM is equal to 67.62 pc/cc. \cite{1}\\
The electrostatic interaction between radio waves and charged particles in the Interstellar Medium creates a delay in the propagation of light, with the delay being a function of radio frequency and the masses of the charged particles or the Dispersion Measure. Lower the frequency, greater is the delay. The delay is given by:
\begin{equation} \label{eq:2}
    \tau (s) = 4.149 \times 10^3 \times DM \times (\nu_1^{-2} - \nu_2^{-2})
\end{equation}
\section{Distance to the Pulsar} \label{section:4}
The distance to the pulsar (S) is given by:
\begin{equation} \label{eq:3}
    S = \frac{DM}{n_e} 
\end{equation}
where $n_e$ is the mean electron density between the pulsar and earth and is equal to 0.23 per cc.\cite{1}, \cite{3}
Hence,
\begin{equation*}
   S = 294\, pc 
\end{equation*}
\section{Dedispersed Time Series}\label{section:5}
We eliminated the frequency-dependent time delays using the DM. To obtain the de-dispersed signal intensity, we changed the time-domain position of all lower frequency channels to align them with the pulse arrival time at the highest channel using Equation \ref{eq:1}, and then added the same for all the channels. The obtained dedispersed time series is given below. The peaks in the above graph are much more significant and easier to recognize among the background noise. This is due to the dedispersion procedure increasing the SNR.
\begin{figure}[!ht]
    \centering
    \includegraphics{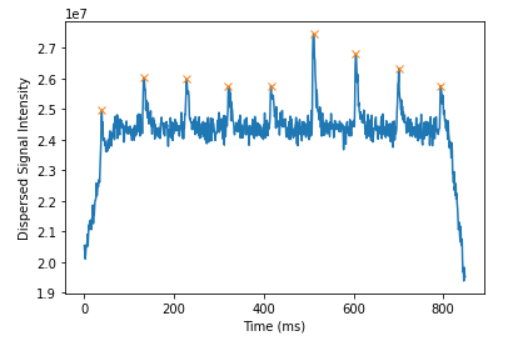}
    \caption{The dedispersed time series.}
    \label{fig:dedispersed}
\end{figure}
\begin{figure}[!ht]
    \centering
    \includegraphics{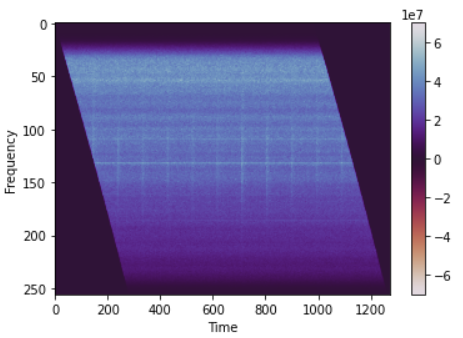}
    \caption{The dynamic spectrum after accounting for the time delay}
    \label{fig:dynamic2}
\end{figure}
\pagebreak
\newpage
\section{Time Period of the Pulsar}
We obtained a series of significant periodic single pulses in the previous section. This enabled us to further calculate the time period of the pulsar.The arrival time of the individual pulse should fit a period-solution, and hence, we used the technique of curve fitting, to estimate the arrival times. The best fit curve is a linear curve as given in the figure below. 
\begin{figure}[!htb]
    \centering
    \includegraphics{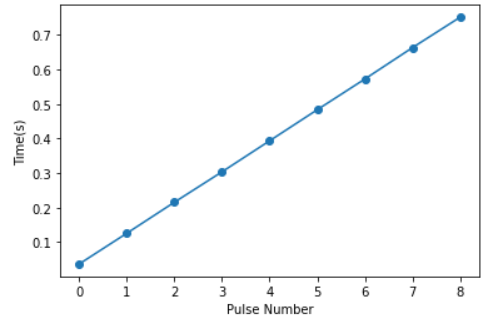}
    \caption{Fitting a linear curve to estimate the arrival times}
    \label{fig:curvefit}
\end{figure}
The arrival time of each pulse has been tabulated below.
\begin{table}[!htb]
\begin{center}
\begin{tabular}{||c |c| c||} 
 \hline
 Pulse number & Arrival time (ms) & Uncertainty(ms)  \\ [0.5ex] 
 \hline\hline
 1 & 88.96388 & 0.33612  \\ 
 \hline
 2 & 178.8742 & 0.610303  \\
 \hline
 3 & 266.8916 & 1.28255  \\
 \hline
 4 & 356.8019 & 0.610303  \\
 \hline
 5 & 446.7122 & 0.610303  \\ 
 \hline
 5 & 535.6761 & 0.33612  \\ 
 \hline
 5 & 624.6761 & 0.3  \\ 
 \hline
\end{tabular}
\caption{Pulse arrival times.}
\label{table:1}
\end{center}
\end{table}
Based on the arrival times as shown in the table above, the time period of the pulsar is 89.3 ms, which on comparison with ATNF Pulsar Catalogue is correct. \\
Finally, based on the time period, we folded the entire time series with the pulsar period to obtain an average profile for the pulsar.
\begin{figure}[!htb]
    \centering
    \includegraphics{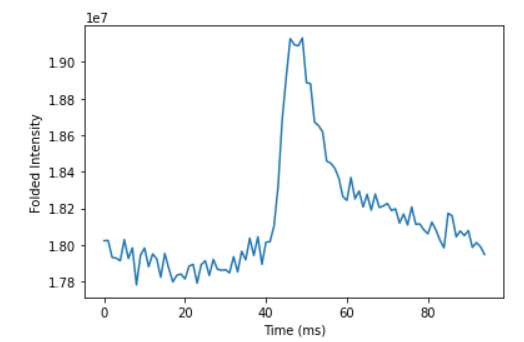}
    \caption{The average profile of the pulsar.}
    \label{fig:profile}
\end{figure}
\newpage
\section{Conclusion}
After implementing various techniques, we managed to calculate important parameters related to the pulsar.We also examined the statistical properties of the raw voltage signal.
\begin{itemize}
    \item We estimated a distance of 294 pc to the pulsar from a Dispersion Measure of 67.62 pc/cc. The distance calculated is in agreement with the value measured using the method of parallax, considering the error margin.\cite{1}
    \item The time period of the pulsar turned out to be 89.3 ms. The calculated time period is in agreement with the currently accepted value.This is more accurate as we used the exact value of Density Measure.\cite{2}
\end{itemize}
An important thing to note is that all the quantitative figures are estimates with some amount of uncertainty. This can be due to uncertainties in other parameters such as the dispersion measure, low exposure time, small dataset etc. Longer observation time would significantly reduce uncertainties in the data. Since the source is a compact object, the visibility should remain constant as a function of time. Hence, the Fourier Transform of brightness distribution should not change with a change in baseline. Over the course of the project, we gained invaluable insights into the working of pulsars and how to decipher information from mere observations.\\ \\
The code used for analysis is available \\ \href{https://colab.research.google.com/drive/18cXAyiXNtvH_WJx7A-bZuNzcBDXm36EX?usp=sharing}{\textcolor{blue}{view code}}.\\ \\
\textit{Acknowledgement:} 
The authors would like to express their gratitude to Dr. Avinash Deshpande, who provided us the raw signal data, and Mr. Devansh Shukla, who helped in the course of the project. HG would like to thank Department of Science and Technology of India for the INSPIRE Scholarship for Higher Education (SHE), (DST/INSPIRE/02/2019/011921).

\end{document}